\documentclass[journal=jacsat,manuscript=article]{achemso}

\usepackage{chemformula} 
\usepackage[T1]{fontenc} 
\usepackage{amssymb,amsmath,mathtools}

\newcommand{\MeijerG}[7]{G \begin{smallmatrix} #1 & #2 \\ #3 & #4 \end{smallmatrix} \left( \begin{smallmatrix} #5 \\ #6 \end{smallmatrix} \middle\vert #7 \right) }

\author{Arnab Barman Ray}
\affiliation{The Institute of Optics, University of Rochester}
\author{Arunabh Mukherjee}
\affiliation{The Institute of Optics, University of Rochester}
\author{Liangyu Qiu}
\affiliation{The Institute of Optics, University of Rochester}
\author{Renee Sailus}
\affiliation{Arizona State University}
\author{Sefaattin Tongay}
\affiliation{Arizona State University}
\author{Anthony Nickolas Vamivakas}
\affiliation{The Institute of Optics, University of Rochester}
\alsoaffiliation{Center for coherence and quantum optics, Department of Physics, University of Rochester}

\email{nick.vamivakas@rochester.edu}
\affiliation{The Institute of Optics, University of Rochester}

\title[An \textsf{achemso} demo]
  {Interplay of trapped species and absence of electron capture in Moir\'{e} heterobilayers}	

\abbreviations{IR,NMR,UV}

\begin{document}

\date{\today}

\begin{abstract}
Moir\'{e} heterobilayers host interlayer excitons in a natural, periodic array of trapping potentials. Recent work has elucidated the structure of the trapped interlayer excitons and the nature of photoluminescence (PL) from trapped and itinerant charged complexes such as interlayer trions in these structures. In this paper, our results serve to add to the understanding of the nature of PL emission and explain its characteristic blueshift with increasing carrier density, along with demonstrating a significant difference between the interlayer exciton-trion conversion efficiency as compared to both localized and itinerant intra-layer species in conventional monolayers. Our results show the absence of optical generation of trions in these materials, which we suggest arises from the highly localized, near sub-nm confinement of trapped species in these Moir\'{e} potentials.

\end{abstract}

\maketitle


\section{Introduction}

Heterostructures, made by stacking monolayers of transition metal dichalcogenides\cite{1,2,3,4,5,6,7} (TMDCs), as well as other materials, such as insulating hexagonal boron nitride(hBN)\cite{8,9}, have attracted significant scientific interest over the last decade. The possibilities of constructing different kinds of condensed matter systems\cite{10,11,12} through the layer degree of freedom and the choice of material have been a cornerstone of research in this direction. In recent years, it was found that bilayers, consisting of two monolayers of either the same (homobilayer) or different (heterobilayer) monolayers can host luminescent long-lived interlayer excitons\cite{13,14}. The longer lifetimes of interlayer excitons in these heterostructures, as opposed to intralayer excitons in monolayers, have made them a material of choice for excitonic applications\cite{15,16}.
More recently, it has been found that two monolayers in contact with each other in a heterobilayer (hBL), develop a periodically varying potential that can trap these interlayer excitons\cite{17,18,19,20}. The structures, called Moiré superlattices, have a period that depends critically on the twist angle between these layers and the mismatch between the lattice period of the two layers themselves\cite{21,22}. These hBLs were shown to host single-photon quantum emitters at the Moiré trapping sites, which also offer a degree of “programmability” through the use of magnetic and electric fields, which can be used to tune the emission wavelengths\cite{23} and polarization\cite{24}.

$\text{MoSe}_2/\text{WSe}_2$ hBLs is one of the most well studied TMDC Moiré systems\cite{25,26}. They are characterized by a large Moiré period (~100 nm) with near $0^\circ$ alignment, owing to their small lattice mismatch between the two monolayer. More recently, isolated trapped species have been observed, trions, bi-excitons, and evidence for higher order charged complexes have emerged\cite{27,28,29}. Beyond isolated emitters, at higher excitation intensities, different species of trions have been identified in this system\cite {30}. Moiré trions are very interesting as they allow a higher degree of nonlinearity in their interactions with light through a phase-space filling effect combined with Moiré localization effects\cite{RN29}. They may have applications for nonlinear photonic devices in quantum computing applications\cite{31,32}. In this regard, a thorough understanding of the physics of Moiré trions is desirable.

The photoluminescence (PL) emission from interlayer excitons in $\text{MoSe}_2/\text{WSe}_2$ Moiré heterobilayers has been studied in some detail, and three broad phases of excited state species were reported from diffusion-based studies - trapped  excitons, itinerant excitons, and electron-hole plasmas at very high excitation intensities equivalent to carrier densities of the order of $10^{13}\,\text{cm}^{-2}$ \cite{4,33}. However, at intermediate carrier densities in the range of $10^{10}-10^{11}\,\text{cm}^{-2}$, when the spectral lines from individual trapped emitters are no longer isolatable due to power broadening and higher densities of trapped emitters, the dynamics between different species are not clear. This state of the system, which we call the quantum ensemble regime\cite{30}, represents emission from a collection of different trapped species across the excitation spot. 

Our results investigate this ensemble, and we observe, by normalizing the PL, that trions, excitons, and bi-excitons exist in this regime and are seen as spectrally separated emission bands. We find a spectral weight transfer across the different species with increasing optical excitation density. The strong correlation of the trion PL with the amount of electrostatically doped electrons in the system indicates an absence of trion-exciton conversion at higher excitation fluxes typical of free and trapped intralayer excitons in monolayers\cite{35,36,37}. We note that the absence of optically generated trions is a feature that distinguishes Moiré systems from conventional monolayers that have been studied so far, with itinerant\cite{34,35} or defect-localized intra-layer excitons\cite{36}. The blue shift observed in the PL at lower excitation intensities has been attributed to the dipolar repulsion between interlayer excitons across Moiré trapping sites\cite{30}. However, our results show that most of the blue-shift that has been observed consistently so far arise primarily from a spectral weight transfer from one species to the other with increasing excitation intensity.

\section{Sample and Methods}
\begin{figure}[t]
\centering
\vspace{0.5cm}
\includegraphics[height=0.7\textwidth,width=0.9\textwidth]{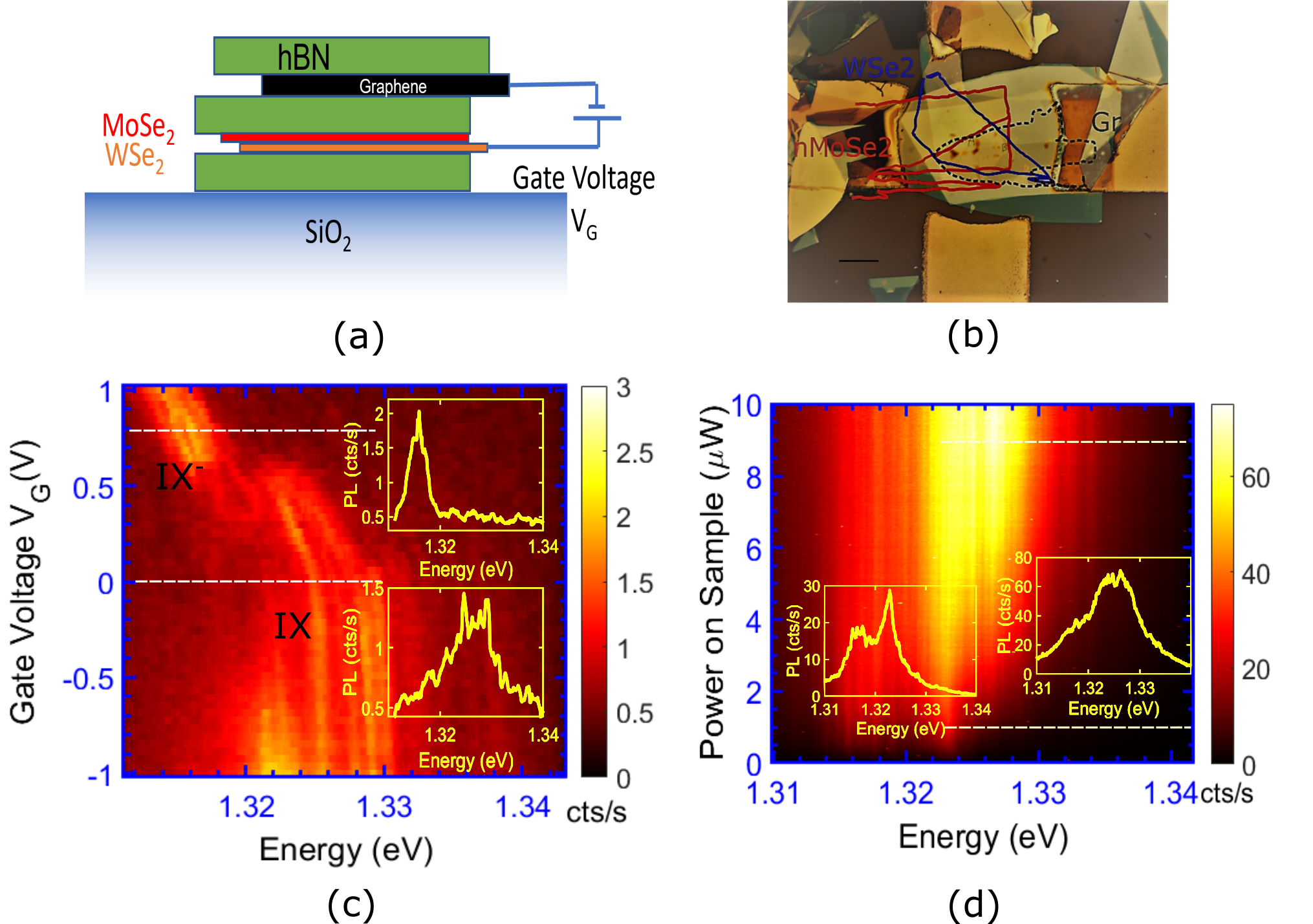}
\caption{(a)A schematic of the kind of devices investigated, (b) optical image of the first sample, the scale bar is 10$\mu$m, (c) PL from interlayer excitons and trions showing a transition with increased doping voltage at an excitation intensity of 20 nW, the insets are a plot of the PL spectra at $V_G = 0.8\,\text{V}$(top) $V_G = 0.0\,\text{V}$(bottom), (d) evolution of the PL with increasing laser intensity, with insets at $1$ $\mu$W (left) and $9$ $\mu$W (right).}
\label{fig1}
\end{figure}

We use a dry transfer technique to assemble a double hBN encapsulated $\text{MoSe}_2-\text{WSe}_2$ single-gated heterostructure with a PC stamp under a microscope. The hBN was obtained from 2D semiconductors. The bulk $\text{MoSe}_2$ was n-type from HQ Graphene, and the WSe2 was prepared at ASU. The monolayers were aligned along precise straight edges, which allowed us to make luminescent hBLs, but without any control on whether the resulting stack would be R-type or H-type\cite{21}, although magnetic measurements in this sample confirm from the observed g-factors, that the sample is R-type\cite{supl}. We use an n-doped monolayer to observe trions without any electrostatic doping. We assemble the hBL on top of a chip with 285 nm of thermally grown $\text{SiO}_2$ on Si. The back gate consists of an FLG in contact with one of the 50 nm gold electrodes. The monolayers are also in contact with another electrode, allowing us to bias the device as a parallel plate capacitor. An accurate voltage source (ANC 300 controller) is used to bias the device. Figure~\ref{fig1}(b) shows an optical micrograph of the device under consideration.

The measurements were carried out using a home-built confocal microscope. A PID stabilized 532 nm DPSS laser is focused into a sub-um diameter spot using a $0.82$ NA objective lens in a closed-cycle cryostat (AttoDry 1000) at 4K equipped with a super-conducting magnet. The PL emitted is collected by the same objective and coupled into a multimode fiber. The collected PL is analyzed using a Princeton Instruments spectrometer (Acton SP-2750i) and an LN2 cooled Pylon CCD camera. We used an intensity calibrator (IntelliCal, Princeton Instruments) to account for the drop in quantum efficiency of the Si CCD camera at near-IR wavelengths.

\section{Results}

First, we investigate the PL from the sample at low carrier densities, when the narrow emission from the individual emitters is well differentiated. The PL we investigate from this sample (R-type) arises from the singlet exciton\cite{br,ec}. Figure~\ref{fig1}(c) shows a voltage sweep, where the voltage is systematically incremented until we see a sharp transition from trapped interlayer excitons to trions. The narrow emission peaks are characteristics of these quantum emitters with individual linewidths $<\, 100 \mu\text{eV}$. We then focus on how the PL evolves with increasing laser intensity in Fig.~\ref{fig1}(d). We note a blue-shift in the observed PL by about ~8 meV. The linewidth of the emission is around 5-6 meV. We note the non-linear evolution of the PL with excitation intensity. For the highest excitation intensities considered in this paper, we estimate a charge carrier density of $~ 5 \times 10^{11}\, \text{cm}^{-2}$, well within the regime where the emission is dominated by moiré trapped excitonic species\cite{33}. We arrive at this estimate by using the steady state equation, $n=\alpha I\tau$ and a carrier lifetime of $\tau\,=\,2\,\text{ns}$\cite{13}, an absorption coefficient\cite{RN12} of $\alpha = 0.10$ at $2.33\,\text{eV}$ in a spot size of area $1\,\mu m^2$.

\begin{figure}[h]
\centering
\vspace{0.5cm}
\includegraphics[height=0.7\textwidth,width=0.9\textwidth]{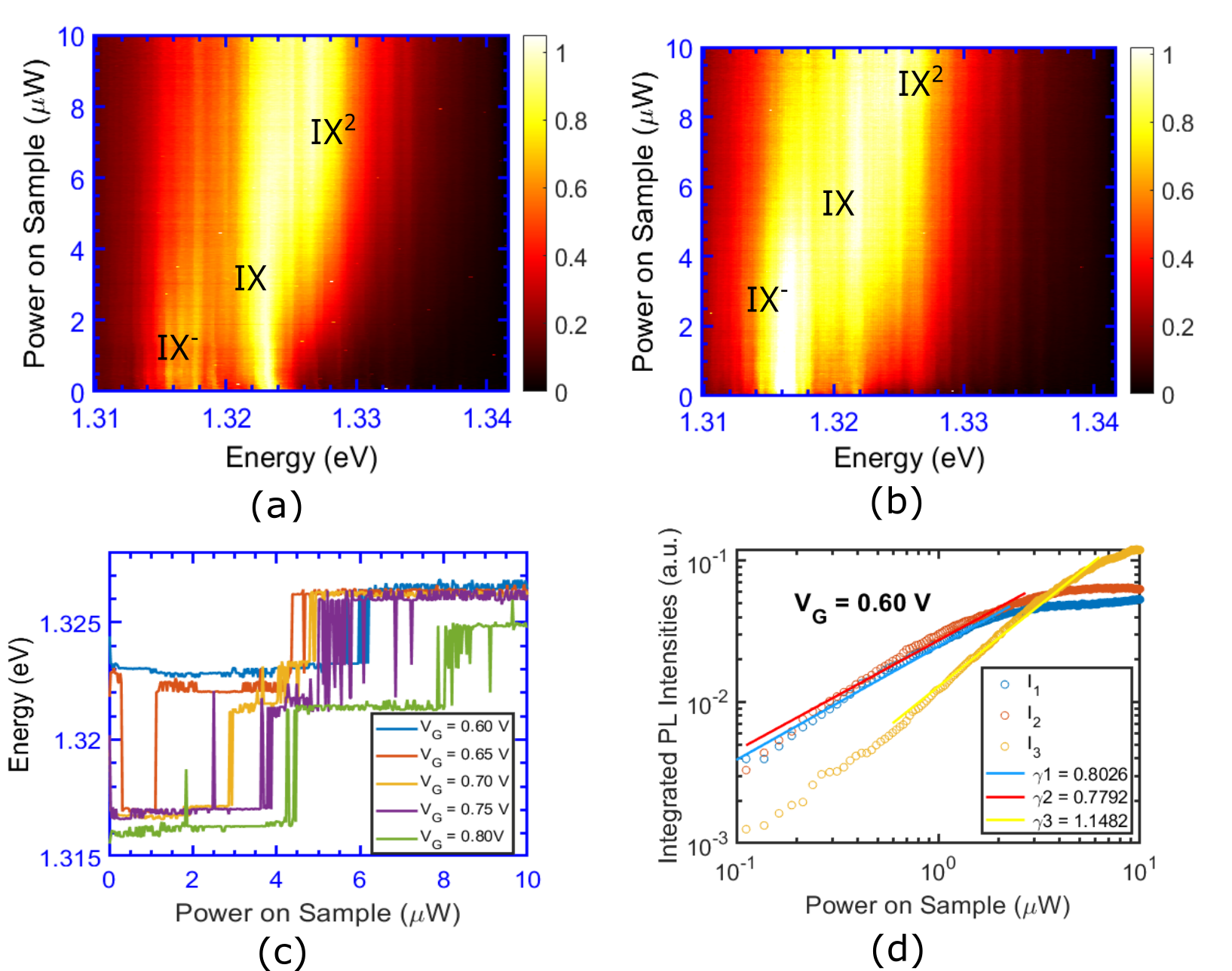}
\caption{ Normalized PL spectra vs power on sample at doping voltages of (a) $V_G = 0.6\, V$, and (b) $V_G = 0.8\, V$,(c) traces of the peak PL intensity with excitation power at different doping voltages, (d) power law fits of the integrated PL intensities of the three species, $I_1$, $I_2$ and $I_3$ correspond to trions, excitons, and bi-excitons respectively.}
\label{fig2}
\end{figure}

We normalize the PL spectra in Fig.~\ref{fig2}(a) to gain further insight. We normalize the PL emission at a given excitation intensity by dividing the spectrum by the observed peak counts. We see that normalizing the PL resolves the emission into three distinct bands. The energy difference between the first two bands is obtained by subtracting the energy difference between the center of the two bands and is found to be around $5-6\, \text{meV}$. The energy difference between the second and third band is around $3-4\,\text{meV}$. We note that these values are consistent with reported values of the binding energy for trion\cite{29,30}, and the additional energy that the trapped bi-exciton exhibits due to intra-trap dipolar repulsion respectively\cite{27}. The binding energies for the trion are much lower than are seen for intra-layer trions due to the effects of localization on the trion wavefunction. Similarly, the spatial proximity of two excitons in the same Moiré trap and the extreme localization causes the bi-exciton to emit at a higher energy.

To confirm the nature of these bands, we fit the bands with separate lorentzians with linewidths of $2\,\text{meV}$ each. The integrated PL intensities obtained from these fits are than plotted as a function of laser power and subjected to a linear fit in Fig.~\ref{fig2}(d). The highest energy band exhibits a growth coefficient of $\gamma_3\,>\,1$. This superlinear growth confirms the bi-excitonic nature of the highest energy band. On the other hand, the doping voltage dependence of the first band confirms that it has a trion origin. The polarity of the applied doping voltages confirms the negatively charged nature of the trions. We note that the small linewidths of each emission band $\sim \, 2\, \text{meV}$ is comparable to the smallest linewidths reported so far, reflecting on the quality of the materials used and the sample quality. Our results show that at higher excitation densities, the PL emission from these samples is predominantly bi-excitonic in origin. Moreover, the differences between the energies of the excitons and bi-excitons, which arise from intra-trap dipolar interactions, can explain most of the blue shift that is seen in the PL from these samples.

We next shift our attention to how the PL evolves with power as a function of the doping density. We trace the peak PL energies at different gate voltages in Fig.~\ref{fig2}(c). We are able to see the shifts in spectral weight from trions to excitons and bi-excitons with increasing laser power. We also notice a DC stark effect, due to the non-zero electric field across the sample. The most exciting feature we observed was that the trion-to-exciton spectral weight jump was a function of the doping density of the sample, and the trion emission dominated the PL spectrum up to higher carrier densities in the sample for larger doping densities.

In monolayers with itinerant or defect-localized excitons, at higher laser fluences, the exciton-trion ratio decreases\cite{34,35,36}. This is due to the prevalence of unbound free carriers. Trions can then form from excitons through electron capture\cite{37} from a sea of unbound charge carriers. This effect tends to increase with increased density of unbound free carriers at higher laser fluences, as the exciton-electron interaction increases\cite{38}. Localization, as seen in the case of excitonic quantum emitters based on defects\cite{36,38}, still allows this exciton-electron interaction. However, our studies reveal the absence of such a mechanism for Moiré localized excitons, suggesting a more robust localization, with the reduced spatial extent and center-of-mass motion\cite{39} resulting in a reduced interaction with the sea of free carriers.

\begin{figure}[h]
\centering
\vspace{0.5cm}
\includegraphics[height=0.6\textwidth,width=1.0\textwidth]{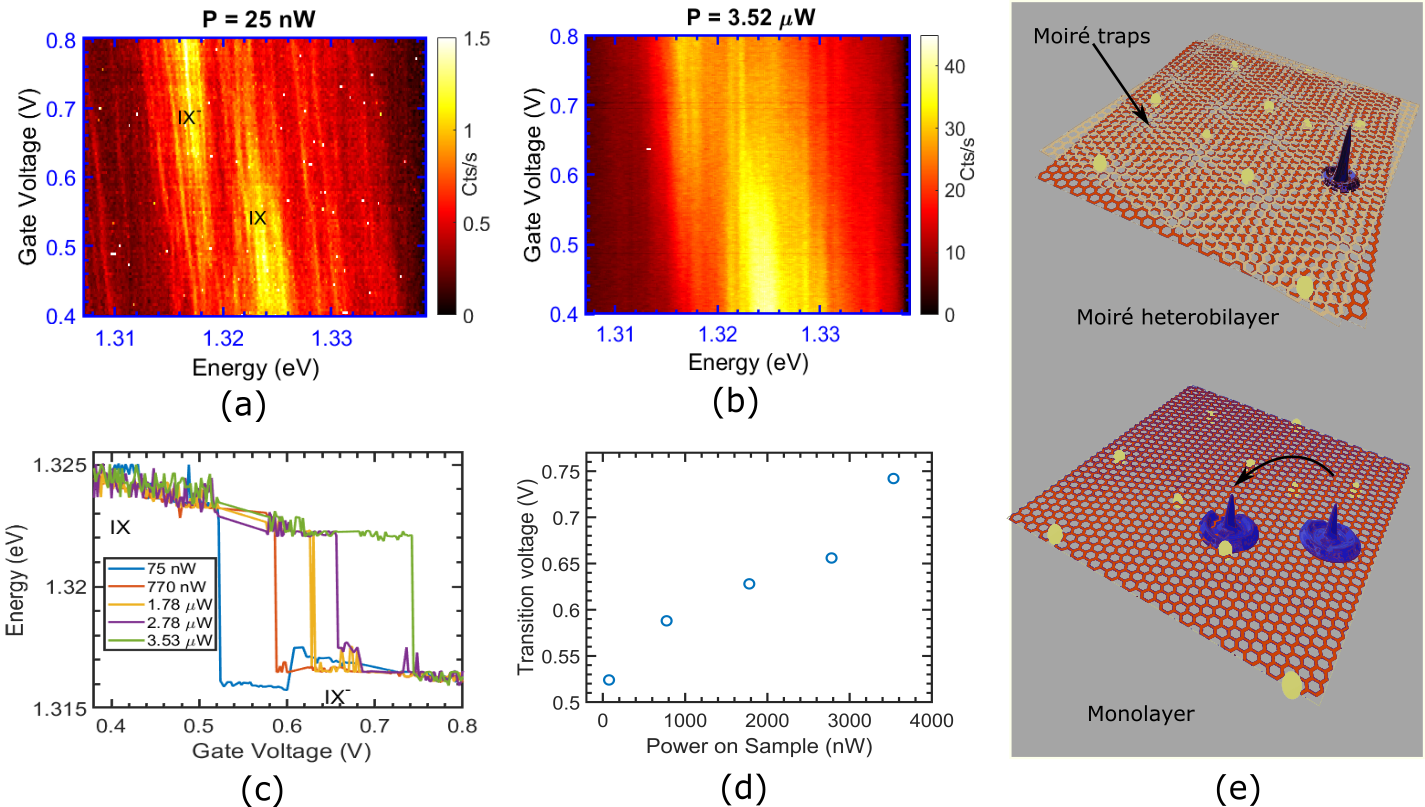}
\caption{ PL spectra with vs doping voltage at excitation pwoers of (a) 25 nW, and (b) 3.62 $\mu$W (c) traces of the peak PL intensity with doping voltage at different excitation powers, (d) transition voltages as a function of power on sample, (e) a cartoon illustrating the difference between highly localized excitons in a Moire trap and an itinerant exciton in a monolayer, yellow spheres are free carriers and the blue hat profiles are representatives of the respective excitonic wavefunctions.}
\label{fig3}
\end{figure}

To understand the interplay between electrostatic doping and laser-created electron-hole carriers, we investigate how the exciton-trion transition for the case of a doping voltage sweep shifts with increasing excitation intensity. In Fig.~\ref{fig3} (a) and (b), we present the PL spectrum in a voltage sweep for two different excitation intensities. We see that the exciton-trion crossover takes place at a larger doping density at higher laser intensities. We plot the peak PL energy as a function of gate voltage for different excitation intensities in Fig.~\ref{fig3} (c); this allows us to see that the transition voltage shifts with increasing laser power. Figure~\ref{fig3} (d) plots the transition voltages as a function of laser power. We see that at higher laser fluences, it takes a larger amount of dopant electrons to shift the dominant optical transition of the system to trions. The linear correlation between the doping voltage and the transition voltage suggests that any trions formed in the system are through the presence of doped electrons sitting at the Moiré traps\cite{RN30}.

To summarize, our results paint the following picture. We start with an array of Moiré traps with a constant density of doped electrons at the trap sites. As we excite the system with a laser, the trion states- the lowest excited states, start filling up. This is when the PL is mostly trion-ic in origin. At higher laser fluences, as all the doped electrons are bound as trions, the absence of electron capture makes it impossible for the population of trions to grow; hence the Moire traps start filling up with excitons. When all the traps are nearly filled, bi-excitonic emission takes over. We note that the presence of charged bi-excitonic species cannot be ruled out; however, further studies are required to mount evidence for their existence.

\section{Conclusions and Future Work}
Our work illustrates the interplay of different species in the PL emission from $\text{MoSe}_2/\text{WSe}_2$ Moiré heterobilayers. Normalizing the PL spectra and tracing its evolution with excitation power shows that most of the observed blueshift in these samples can be attributed to the spectral weight transfer in the PL across different species and hence to intra-trap dipole-dipole interactions as opposed to interactions across different Moire sites. We find that at higher laser fluences (carrier densities $< \, 10^{12} \, \text{cm}^{-2}$), the PL is dominated by bi-excitonic emission. By analyzing the trion-exciton spectral weight, we also show that Moiré localization leads to an absence of electron capture in these systems, which hints at a qualitative difference in the strength of localization over conventional defect-based quantum emitters in monolayers.
Investigating the time dynamics of trion formation and a rigorous rate equation analysis of these systems may shed more light on the exciton-electron interactions. Two-dimensional Fourier spectroscopy may also reveal how different trapped species are coupled to each other by analyzing the cross-couplings and may shed more light on the excited state landscape in these systems.

\section{Acknowledgments}
	
This work was supported by
AFOSR FA9550-19-1-0074 from the Cornell Center for Materials Research with funding from the NSF MRSEC program (DMR-1719875). S.T acknowledges NSF CMMI 2129412 and NSF DMR 2111812.


	\bibliography{biblio}

\providecommand{\latin}[1]{#1}
\makeatletter
\providecommand{\doi}
  {\begingroup\let\do\@makeother\dospecials
  \catcode`\{=1 \catcode`\}=2 \doi@aux}
\providecommand{\doi@aux}[1]{\endgroup\texttt{#1}}
\makeatother
\providecommand*\mcitethebibliography{\thebibliography}
\csname @ifundefined\endcsname{endmcitethebibliography}
  {\let\endmcitethebibliography\endthebibliography}{}
\begin{mcitethebibliography}{46}
\providecommand*\natexlab[1]{#1}
\providecommand*\mciteSetBstSublistMode[1]{}
\providecommand*\mciteSetBstMaxWidthForm[2]{}
\providecommand*\mciteBstWouldAddEndPuncttrue
  {\def\EndOfBibitem{\unskip.}}
\providecommand*\mciteBstWouldAddEndPunctfalse
  {\let\EndOfBibitem\relax}
\providecommand*\mciteSetBstMidEndSepPunct[3]{}
\providecommand*\mciteSetBstSublistLabelBeginEnd[3]{}
\providecommand*\EndOfBibitem{}
\mciteSetBstSublistMode{f}
\mciteSetBstMaxWidthForm{subitem}{(\alph{mcitesubitemcount})}
\mciteSetBstSublistLabelBeginEnd
  {\mcitemaxwidthsubitemform\space}
  {\relax}
  {\relax}

\bibitem[He \latin{et~al.}(2014)He, Kumar, Zhao, Wang, Mak, Zhao, and Shan]{1}
He,~K.; Kumar,~N.; Zhao,~L.; Wang,~Z.; Mak,~K.~F.; Zhao,~H.; Shan,~J. Tightly
  bound excitons in {MonolayerWSe2}. \emph{Phys. Rev. Lett.} \textbf{2014},
  \emph{113}\relax
\mciteBstWouldAddEndPuncttrue
\mciteSetBstMidEndSepPunct{\mcitedefaultmidpunct}
{\mcitedefaultendpunct}{\mcitedefaultseppunct}\relax
\EndOfBibitem
\bibitem[Mak \latin{et~al.}(2013)Mak, He, Lee, Lee, Hone, Heinz, and Shan]{2}
Mak,~K.~F.; He,~K.; Lee,~C.; Lee,~G.~H.; Hone,~J.; Heinz,~T.~F.; Shan,~J.
  Tightly bound trions in monolayer {MoS2}. \emph{Nat. Mater.} \textbf{2013},
  \emph{12}, 207--211\relax
\mciteBstWouldAddEndPuncttrue
\mciteSetBstMidEndSepPunct{\mcitedefaultmidpunct}
{\mcitedefaultendpunct}{\mcitedefaultseppunct}\relax
\EndOfBibitem
\bibitem[Jones \latin{et~al.}(2013)Jones, Yu, Ghimire, Wu, Aivazian, Ross,
  Zhao, Yan, Mandrus, Xiao, Yao, and Xu]{3}
Jones,~A.~M.; Yu,~H.; Ghimire,~N.~J.; Wu,~S.; Aivazian,~G.; Ross,~J.~S.;
  Zhao,~B.; Yan,~J.; Mandrus,~D.~G.; Xiao,~D.; Yao,~W.; Xu,~X. Optical
  generation of excitonic valley coherence in monolayer {WSe2}. \emph{Nat.
  Nanotechnol.} \textbf{2013}, \emph{8}, 634--638\relax
\mciteBstWouldAddEndPuncttrue
\mciteSetBstMidEndSepPunct{\mcitedefaultmidpunct}
{\mcitedefaultendpunct}{\mcitedefaultseppunct}\relax
\EndOfBibitem
\bibitem[Barman~Ray \latin{et~al.}(2022)Barman~Ray, Liang, and Vamivakas]{4}
Barman~Ray,~A.; Liang,~K.; Vamivakas,~A.~N. Valley engineering electron-hole
  liquids in transition metal dichalcogenide monolayers. \emph{Phys. Rev. B}
  \textbf{2022}, \emph{106}, 045206\relax
\mciteBstWouldAddEndPuncttrue
\mciteSetBstMidEndSepPunct{\mcitedefaultmidpunct}
{\mcitedefaultendpunct}{\mcitedefaultseppunct}\relax
\EndOfBibitem
\bibitem[Splendiani \latin{et~al.}(2010)Splendiani, Sun, Zhang, Li, Kim, Chim,
  Galli, and Wang]{5}
Splendiani,~A.; Sun,~L.; Zhang,~Y.; Li,~T.; Kim,~J.; Chim,~C.-Y.; Galli,~G.;
  Wang,~F. Emerging Photoluminescence in Monolayer MoS2. \emph{Nano Letters}
  \textbf{2010}, \emph{10}, 1271--1275\relax
\mciteBstWouldAddEndPuncttrue
\mciteSetBstMidEndSepPunct{\mcitedefaultmidpunct}
{\mcitedefaultendpunct}{\mcitedefaultseppunct}\relax
\EndOfBibitem
\bibitem[Castellanos-Gomez(2016)]{6}
Castellanos-Gomez,~A. Why all the fuss about 2D semiconductors? \emph{Nature
  Photonics} \textbf{2016}, \emph{10}, 202--204\relax
\mciteBstWouldAddEndPuncttrue
\mciteSetBstMidEndSepPunct{\mcitedefaultmidpunct}
{\mcitedefaultendpunct}{\mcitedefaultseppunct}\relax
\EndOfBibitem
\bibitem[Mak and Shan(2016)Mak, and Shan]{7}
Mak,~K.~F.; Shan,~J. Photonics and optoelectronics of 2D semiconductor
  transition metal dichalcogenides. \emph{Nature Photonics} \textbf{2016},
  \emph{10}, 216--226\relax
\mciteBstWouldAddEndPuncttrue
\mciteSetBstMidEndSepPunct{\mcitedefaultmidpunct}
{\mcitedefaultendpunct}{\mcitedefaultseppunct}\relax
\EndOfBibitem
\bibitem[Cadiz \latin{et~al.}(2017)Cadiz, Courtade, Robert, Wang, Shen, Cai,
  Taniguchi, Watanabe, Carrere, Lagarde, Manca, Amand, Renucci, Tongay, Marie,
  and Urbaszek]{8}
Cadiz,~F. \latin{et~al.}  Excitonic Linewidth Approaching the Homogeneous Limit
  in {MoS2} -Based van der Waals Heterostructures. \emph{Phys. Rev. X.}
  \textbf{2017}, \emph{7}\relax
\mciteBstWouldAddEndPuncttrue
\mciteSetBstMidEndSepPunct{\mcitedefaultmidpunct}
{\mcitedefaultendpunct}{\mcitedefaultseppunct}\relax
\EndOfBibitem
\bibitem[Taniguchi and Watanabe(2007)Taniguchi, and Watanabe]{9}
Taniguchi,~T.; Watanabe,~K. Synthesis of high-purity boron nitride single
  crystals under high pressure by using Ba–BN solvent. \emph{Journal of
  Crystal Growth} \textbf{2007}, \emph{303}, 525--529\relax
\mciteBstWouldAddEndPuncttrue
\mciteSetBstMidEndSepPunct{\mcitedefaultmidpunct}
{\mcitedefaultendpunct}{\mcitedefaultseppunct}\relax
\EndOfBibitem
\bibitem[Gu \latin{et~al.}(2022)Gu, Ma, Liu, Watanabe, Taniguchi, Hone, Shan,
  and Mak]{10}
Gu,~J.; Ma,~L.; Liu,~S.; Watanabe,~K.; Taniguchi,~T.; Hone,~J.~C.; Shan,~J.;
  Mak,~K.~F. Dipolar excitonic insulator in a moiré lattice. \emph{Nature
  Physics} \textbf{2022}, \emph{18}, 395--400\relax
\mciteBstWouldAddEndPuncttrue
\mciteSetBstMidEndSepPunct{\mcitedefaultmidpunct}
{\mcitedefaultendpunct}{\mcitedefaultseppunct}\relax
\EndOfBibitem
\bibitem[Chen \latin{et~al.}(2022)Chen, Lian, Huang, Su, Rashetnia, Ma, Yan,
  Blei, Xiang, Taniguchi, Watanabe, Tongay, Smirnov, Wang, Zhang, Cui, and
  Shi]{11}
Chen,~D. \latin{et~al.}  Excitonic insulator in a heterojunction moiré
  superlattice. \emph{Nature Physics} \textbf{2022}, \emph{18},
  1171--1176\relax
\mciteBstWouldAddEndPuncttrue
\mciteSetBstMidEndSepPunct{\mcitedefaultmidpunct}
{\mcitedefaultendpunct}{\mcitedefaultseppunct}\relax
\EndOfBibitem
\bibitem[Tang \latin{et~al.}(2020)Tang, Li, Li, Xu, Liu, Barmak, Watanabe,
  Taniguchi, MacDonald, Shan, and Mak]{12}
Tang,~Y.; Li,~L.; Li,~T.; Xu,~Y.; Liu,~S.; Barmak,~K.; Watanabe,~K.;
  Taniguchi,~T.; MacDonald,~A.~H.; Shan,~J.; Mak,~K.~F. Simulation of Hubbard
  model physics in WSe2/WS2 moiré superlattices. \emph{Nature} \textbf{2020},
  \emph{579}, 353--358\relax
\mciteBstWouldAddEndPuncttrue
\mciteSetBstMidEndSepPunct{\mcitedefaultmidpunct}
{\mcitedefaultendpunct}{\mcitedefaultseppunct}\relax
\EndOfBibitem
\bibitem[Rivera \latin{et~al.}(2015)Rivera, Schaibley, Jones, Ross, Wu,
  Aivazian, Klement, Seyler, Clark, Ghimire, Yan, Mandrus, Yao, and Xu]{13}
Rivera,~P.; Schaibley,~J.~R.; Jones,~A.~M.; Ross,~J.~S.; Wu,~S.; Aivazian,~G.;
  Klement,~P.; Seyler,~K.; Clark,~G.; Ghimire,~N.~J.; Yan,~J.; Mandrus,~D.~G.;
  Yao,~W.; Xu,~X. Observation of long-lived interlayer excitons in monolayer
  {MoSe2--WSe2} heterostructures. \emph{Nat. Commun.} \textbf{2015},
  \emph{6}\relax
\mciteBstWouldAddEndPuncttrue
\mciteSetBstMidEndSepPunct{\mcitedefaultmidpunct}
{\mcitedefaultendpunct}{\mcitedefaultseppunct}\relax
\EndOfBibitem
\bibitem[Miller \latin{et~al.}(2017)Miller, Steinhoff, Pano, Klein, Jahnke,
  Holleitner, and Wurstbauer]{14}
Miller,~B.; Steinhoff,~A.; Pano,~B.; Klein,~J.; Jahnke,~F.; Holleitner,~A.;
  Wurstbauer,~U. Long-lived direct and indirect interlayer excitons in van der
  Waals heterostructures. \emph{Nano Lett.} \textbf{2017}, \emph{17},
  5229--5237\relax
\mciteBstWouldAddEndPuncttrue
\mciteSetBstMidEndSepPunct{\mcitedefaultmidpunct}
{\mcitedefaultendpunct}{\mcitedefaultseppunct}\relax
\EndOfBibitem
\bibitem[Unuchek \latin{et~al.}(2018)Unuchek, Ciarrocchi, Avsar, Watanabe,
  Taniguchi, and Kis]{15}
Unuchek,~D.; Ciarrocchi,~A.; Avsar,~A.; Watanabe,~K.; Taniguchi,~T.; Kis,~A.
  Room-temperature electrical control of exciton flux in a van der Waals
  heterostructure. \emph{Nature} \textbf{2018}, \emph{560}, 340--344\relax
\mciteBstWouldAddEndPuncttrue
\mciteSetBstMidEndSepPunct{\mcitedefaultmidpunct}
{\mcitedefaultendpunct}{\mcitedefaultseppunct}\relax
\EndOfBibitem
\bibitem[Sun \latin{et~al.}(2022)Sun, Ciarrocchi, Tagarelli, Gonzalez~Marin,
  Watanabe, Taniguchi, and Kis]{16}
Sun,~Z.; Ciarrocchi,~A.; Tagarelli,~F.; Gonzalez~Marin,~J.~F.; Watanabe,~K.;
  Taniguchi,~T.; Kis,~A. Excitonic transport driven by repulsive dipolar
  interaction in a van der Waals heterostructure. \emph{Nat. Photonics}
  \textbf{2022}, \emph{16}, 79--85\relax
\mciteBstWouldAddEndPuncttrue
\mciteSetBstMidEndSepPunct{\mcitedefaultmidpunct}
{\mcitedefaultendpunct}{\mcitedefaultseppunct}\relax
\EndOfBibitem
\bibitem[Jin \latin{et~al.}(2019)Jin, Regan, Yan, Iqbal Bakti~Utama, Wang,
  Zhao, Qin, Yang, Zheng, Shi, Watanabe, Taniguchi, Tongay, Zettl, and
  Wang]{17}
Jin,~C.; Regan,~E.~C.; Yan,~A.; Iqbal Bakti~Utama,~M.; Wang,~D.; Zhao,~S.;
  Qin,~Y.; Yang,~S.; Zheng,~Z.; Shi,~S.; Watanabe,~K.; Taniguchi,~T.;
  Tongay,~S.; Zettl,~A.; Wang,~F. Observation of moir{\'e} excitons in
  {WSe2/WS2} heterostructure superlattices. \emph{Nature} \textbf{2019},
  \emph{567}, 76--80\relax
\mciteBstWouldAddEndPuncttrue
\mciteSetBstMidEndSepPunct{\mcitedefaultmidpunct}
{\mcitedefaultendpunct}{\mcitedefaultseppunct}\relax
\EndOfBibitem
\bibitem[F{\"o}rg \latin{et~al.}(2021)F{\"o}rg, Baimuratov, Kruchinin, Vovk,
  Scherzer, F{\"o}rste, Funk, Watanabe, Taniguchi, and H{\"o}gele]{18}
F{\"o}rg,~M.; Baimuratov,~A.~S.; Kruchinin,~S.~Y.; Vovk,~I.~A.; Scherzer,~J.;
  F{\"o}rste,~J.; Funk,~V.; Watanabe,~K.; Taniguchi,~T.; H{\"o}gele,~A.
  Moir{\'e} excitons in {MoSe2-WSe2} heterobilayers and heterotrilayers.
  \emph{Nat. Commun.} \textbf{2021}, \emph{12}\relax
\mciteBstWouldAddEndPuncttrue
\mciteSetBstMidEndSepPunct{\mcitedefaultmidpunct}
{\mcitedefaultendpunct}{\mcitedefaultseppunct}\relax
\EndOfBibitem
\bibitem[Yu \latin{et~al.}(2017)Yu, Liu, Tang, Xu, and Yao]{19}
Yu,~H.; Liu,~G.-B.; Tang,~J.; Xu,~X.; Yao,~W. Moir{\'e} excitons: From
  programmable quantum emitter arrays to spin-orbit--coupled artificial
  lattices. \emph{Sci. Adv.} \textbf{2017}, \emph{3}\relax
\mciteBstWouldAddEndPuncttrue
\mciteSetBstMidEndSepPunct{\mcitedefaultmidpunct}
{\mcitedefaultendpunct}{\mcitedefaultseppunct}\relax
\EndOfBibitem
\bibitem[Seyler \latin{et~al.}(2019)Seyler, Rivera, Yu, Wilson, Ray, Mandrus,
  Yan, Yao, and Xu]{20}
Seyler,~K.~L.; Rivera,~P.; Yu,~H.; Wilson,~N.~P.; Ray,~E.~L.; Mandrus,~D.~G.;
  Yan,~J.; Yao,~W.; Xu,~X. Signatures of moir{\'e}-trapped valley excitons in
  {MoSe2/WSe2} heterobilayers. \emph{Nature} \textbf{2019}, \emph{567},
  66--70\relax
\mciteBstWouldAddEndPuncttrue
\mciteSetBstMidEndSepPunct{\mcitedefaultmidpunct}
{\mcitedefaultendpunct}{\mcitedefaultseppunct}\relax
\EndOfBibitem
\bibitem[Mak and Shan(2022)Mak, and Shan]{21}
Mak,~K.~F.; Shan,~J. Semiconductor moir{\'e} materials. \emph{Nat.
  Nanotechnol.} \textbf{2022}, \emph{17}, 686--695\relax
\mciteBstWouldAddEndPuncttrue
\mciteSetBstMidEndSepPunct{\mcitedefaultmidpunct}
{\mcitedefaultendpunct}{\mcitedefaultseppunct}\relax
\EndOfBibitem
\bibitem[Liu \latin{et~al.}(2021)Liu, Zeng, Yu, Zhong, Li, Zhang, Liu, Wang,
  Pan, and Duan]{22}
Liu,~Y.; Zeng,~C.; Yu,~J.; Zhong,~J.; Li,~B.; Zhang,~Z.; Liu,~Z.; Wang,~Z.~M.;
  Pan,~A.; Duan,~X. Moir{\'e} superlattices and related moir{\'e} excitons in
  twisted van der Waals heterostructures. \emph{Chem. Soc. Rev.} \textbf{2021},
  \emph{50}, 6401--6422\relax
\mciteBstWouldAddEndPuncttrue
\mciteSetBstMidEndSepPunct{\mcitedefaultmidpunct}
{\mcitedefaultendpunct}{\mcitedefaultseppunct}\relax
\EndOfBibitem
\bibitem[Baek \latin{et~al.}(2020)Baek, Brotons-Gisbert, Koong, Campbell,
  Rambach, Watanabe, Taniguchi, and Gerardot]{23}
Baek,~H.; Brotons-Gisbert,~M.; Koong,~Z.~X.; Campbell,~A.; Rambach,~M.;
  Watanabe,~K.; Taniguchi,~T.; Gerardot,~B.~D. Highly energy-tunable quantum
  light from moir{\'e}-trapped excitons. \emph{Sci. Adv.} \textbf{2020},
  \emph{6}\relax
\mciteBstWouldAddEndPuncttrue
\mciteSetBstMidEndSepPunct{\mcitedefaultmidpunct}
{\mcitedefaultendpunct}{\mcitedefaultseppunct}\relax
\EndOfBibitem
\bibitem[Ciarrocchi \latin{et~al.}(2019)Ciarrocchi, Unuchek, Avsar, Watanabe,
  Taniguchi, and Kis]{24}
Ciarrocchi,~A.; Unuchek,~D.; Avsar,~A.; Watanabe,~K.; Taniguchi,~T.; Kis,~A.
  Polarization switching and electrical control of interlayer excitons in
  two-dimensional van der Waals heterostructures. \emph{Nat. Photonics}
  \textbf{2019}, \emph{13}, 131--136\relax
\mciteBstWouldAddEndPuncttrue
\mciteSetBstMidEndSepPunct{\mcitedefaultmidpunct}
{\mcitedefaultendpunct}{\mcitedefaultseppunct}\relax
\EndOfBibitem
\bibitem[Brem \latin{et~al.}(2020)Brem, Linderälv, Erhart, and Malic]{25}
Brem,~S.; Linderälv,~C.; Erhart,~P.; Malic,~E. Tunable Phases of Moiré
  Excitons in van der Waals Heterostructures. \emph{Nano Letters}
  \textbf{2020}, \emph{20}, 8534--8540\relax
\mciteBstWouldAddEndPuncttrue
\mciteSetBstMidEndSepPunct{\mcitedefaultmidpunct}
{\mcitedefaultendpunct}{\mcitedefaultseppunct}\relax
\EndOfBibitem
\bibitem[Sigl \latin{et~al.}(2022)Sigl, Troue, Katzer, Selig, Sigger, Kiemle,
  Brotons-Gisbert, Watanabe, Taniguchi, Gerardot, Knorr, Wurstbauer, and
  Holleitner]{26}
Sigl,~L.; Troue,~M.; Katzer,~M.; Selig,~M.; Sigger,~F.; Kiemle,~J.;
  Brotons-Gisbert,~M.; Watanabe,~K.; Taniguchi,~T.; Gerardot,~B.~D.; Knorr,~A.;
  Wurstbauer,~U.; Holleitner,~A.~W. Optical dipole orientation of interlayer
  excitons in ${\mathrm{MoSe}}_{2}\ensuremath{-}{\mathrm{WSe}}_{2}$
  heterostacks. \emph{Phys. Rev. B} \textbf{2022}, \emph{105}, 035417\relax
\mciteBstWouldAddEndPuncttrue
\mciteSetBstMidEndSepPunct{\mcitedefaultmidpunct}
{\mcitedefaultendpunct}{\mcitedefaultseppunct}\relax
\EndOfBibitem
\bibitem[Li \latin{et~al.}(2020)Li, Lu, Dubey, Devenica, and Srivastava]{27}
Li,~W.; Lu,~X.; Dubey,~S.; Devenica,~L.; Srivastava,~A. Dipolar interactions
  between localized interlayer excitons in van der Waals heterostructures.
  \emph{Nat. Mater.} \textbf{2020}, \emph{19}, 624--629\relax
\mciteBstWouldAddEndPuncttrue
\mciteSetBstMidEndSepPunct{\mcitedefaultmidpunct}
{\mcitedefaultendpunct}{\mcitedefaultseppunct}\relax
\EndOfBibitem
\bibitem[Liu \latin{et~al.}(2021)Liu, Barré, van Baren, Wilson, Taniguchi,
  Watanabe, Cui, Gabor, Heinz, Chang, and Lui]{28}
Liu,~E.; Barré,~E.; van Baren,~J.; Wilson,~M.; Taniguchi,~T.; Watanabe,~K.;
  Cui,~Y.-T.; Gabor,~N.~M.; Heinz,~T.~F.; Chang,~Y.-C.; Lui,~C.~H. Signatures
  of moiré trions in WSe2/MoSe2 heterobilayers. \emph{Nature} \textbf{2021},
  \emph{594}, 46--50\relax
\mciteBstWouldAddEndPuncttrue
\mciteSetBstMidEndSepPunct{\mcitedefaultmidpunct}
{\mcitedefaultendpunct}{\mcitedefaultseppunct}\relax
\EndOfBibitem
\bibitem[Wang \latin{et~al.}(2021)Wang, Zhu, Seyler, Rivera, Zheng, Wang, He,
  Taniguchi, Watanabe, Yan, Mandrus, Gamelin, Yao, and Xu]{29}
Wang,~X.; Zhu,~J.; Seyler,~K.~L.; Rivera,~P.; Zheng,~H.; Wang,~Y.; He,~M.;
  Taniguchi,~T.; Watanabe,~K.; Yan,~J.; Mandrus,~D.~G.; Gamelin,~D.~R.;
  Yao,~W.; Xu,~X. Moiré trions in MoSe2/WSe2 heterobilayers. \emph{Nature
  Nanotechnology} \textbf{2021}, \emph{16}, 1208--1213\relax
\mciteBstWouldAddEndPuncttrue
\mciteSetBstMidEndSepPunct{\mcitedefaultmidpunct}
{\mcitedefaultendpunct}{\mcitedefaultseppunct}\relax
\EndOfBibitem
\bibitem[Brotons-Gisbert \latin{et~al.}(2021)Brotons-Gisbert, Baek, Campbell,
  Watanabe, Taniguchi, and Gerardot]{30}
Brotons-Gisbert,~M.; Baek,~H.; Campbell,~A.; Watanabe,~K.; Taniguchi,~T.;
  Gerardot,~B.~D. Moir\'e-Trapped Interlayer Trions in a Charge-Tunable
  ${\mathrm{WSe}}_{2}/{\mathrm{MoSe}}_{2}$ Heterobilayer. \emph{Phys. Rev. X}
  \textbf{2021}, \emph{11}, 031033\relax
\mciteBstWouldAddEndPuncttrue
\mciteSetBstMidEndSepPunct{\mcitedefaultmidpunct}
{\mcitedefaultendpunct}{\mcitedefaultseppunct}\relax
\EndOfBibitem
\bibitem[Zhang \latin{et~al.}(2021)Zhang, Wu, Hou, Zhang, Chou, Watanabe,
  Taniguchi, Forrest, and Deng]{RN29}
Zhang,~L.; Wu,~F.; Hou,~S.; Zhang,~Z.; Chou,~Y.-H.; Watanabe,~K.;
  Taniguchi,~T.; Forrest,~S.~R.; Deng,~H. Van der Waals heterostructure
  polaritons with moiré-induced nonlinearity. \emph{Nature} \textbf{2021},
  \emph{591}, 61--65\relax
\mciteBstWouldAddEndPuncttrue
\mciteSetBstMidEndSepPunct{\mcitedefaultmidpunct}
{\mcitedefaultendpunct}{\mcitedefaultseppunct}\relax
\EndOfBibitem
\bibitem[Emmanuele \latin{et~al.}(2020)Emmanuele, Sich, Kyriienko,
  Shahnazaryan, Withers, Catanzaro, Walker, Benimetskiy, Skolnick,
  Tartakovskii, Shelykh, and Krizhanovskii]{31}
Emmanuele,~R. P.~A.; Sich,~M.; Kyriienko,~O.; Shahnazaryan,~V.; Withers,~F.;
  Catanzaro,~A.; Walker,~P.~M.; Benimetskiy,~F.~A.; Skolnick,~M.~S.;
  Tartakovskii,~A.~I.; Shelykh,~I.~A.; Krizhanovskii,~D.~N. Highly nonlinear
  trion-polaritons in a monolayer semiconductor. \emph{Nature Communications}
  \textbf{2020}, \emph{11}, 3589\relax
\mciteBstWouldAddEndPuncttrue
\mciteSetBstMidEndSepPunct{\mcitedefaultmidpunct}
{\mcitedefaultendpunct}{\mcitedefaultseppunct}\relax
\EndOfBibitem
\bibitem[Kyriienko \latin{et~al.}(2020)Kyriienko, Krizhanovskii, and
  Shelykh]{32}
Kyriienko,~O.; Krizhanovskii,~D.~N.; Shelykh,~I.~A. Nonlinear Quantum Optics
  with Trion Polaritons in 2D Monolayers: Conventional and Unconventional
  Photon Blockade. \emph{Phys. Rev. Lett.} \textbf{2020}, \emph{125},
  197402\relax
\mciteBstWouldAddEndPuncttrue
\mciteSetBstMidEndSepPunct{\mcitedefaultmidpunct}
{\mcitedefaultendpunct}{\mcitedefaultseppunct}\relax
\EndOfBibitem
\bibitem[Wang \latin{et~al.}(2021)Wang, Shi, Shih, Zhou, Wu, Bai, Rhodes,
  Barmak, Hone, Dean, and Zhu]{33}
Wang,~J.; Shi,~Q.; Shih,~E.-M.; Zhou,~L.; Wu,~W.; Bai,~Y.; Rhodes,~D.;
  Barmak,~K.; Hone,~J.; Dean,~C.~R.; Zhu,~X.-Y. Diffusivity Reveals Three
  Distinct Phases of Interlayer Excitons in
  ${\mathrm{MoSe}}_{2}/{\mathrm{WSe}}_{2}$ Heterobilayers. \emph{Phys. Rev.
  Lett.} \textbf{2021}, \emph{126}, 106804\relax
\mciteBstWouldAddEndPuncttrue
\mciteSetBstMidEndSepPunct{\mcitedefaultmidpunct}
{\mcitedefaultendpunct}{\mcitedefaultseppunct}\relax
\EndOfBibitem
\bibitem[Kesarwani \latin{et~al.}(2022)Kesarwani, Simbulan, Huang, Chiang, Yeh,
  Lan, and Lu]{35}
Kesarwani,~R.; Simbulan,~K.~B.; Huang,~T.-D.; Chiang,~Y.-F.; Yeh,~N.-C.;
  Lan,~Y.-W.; Lu,~T.-H. Control of trion-to-exciton conversion in monolayer
  WS<sub>2</sub> by orbital angular momentum of light. \emph{Science Advances}
  \textbf{2022}, \emph{8}, eabm0100\relax
\mciteBstWouldAddEndPuncttrue
\mciteSetBstMidEndSepPunct{\mcitedefaultmidpunct}
{\mcitedefaultendpunct}{\mcitedefaultseppunct}\relax
\EndOfBibitem
\bibitem[Chakraborty \latin{et~al.}(2018)Chakraborty, Qiu, Konthasinghe,
  Mukherjee, Dhara, and Vamivakas]{36}
Chakraborty,~C.; Qiu,~L.; Konthasinghe,~K.; Mukherjee,~A.; Dhara,~S.;
  Vamivakas,~N. 3D Localized Trions in Monolayer WSe2 in a Charge Tunable van
  der Waals Heterostructure. \emph{Nano Letters} \textbf{2018}, \emph{18},
  2859--2863\relax
\mciteBstWouldAddEndPuncttrue
\mciteSetBstMidEndSepPunct{\mcitedefaultmidpunct}
{\mcitedefaultendpunct}{\mcitedefaultseppunct}\relax
\EndOfBibitem
\bibitem[Portella-Oberli \latin{et~al.}(2009)Portella-Oberli, Berney, Kappei,
  Morier-Genoud, Szczytko, and Deveaud-Pl\'edran]{37}
Portella-Oberli,~M.~T.; Berney,~J.; Kappei,~L.; Morier-Genoud,~F.;
  Szczytko,~J.; Deveaud-Pl\'edran,~B. Dynamics of Trion Formation in
  ${\mathrm{In}}_{x}{\mathrm{Ga}}_{1\ensuremath{-}x}\mathrm{As}$ Quantum Wells.
  \emph{Phys. Rev. Lett.} \textbf{2009}, \emph{102}, 096402\relax
\mciteBstWouldAddEndPuncttrue
\mciteSetBstMidEndSepPunct{\mcitedefaultmidpunct}
{\mcitedefaultendpunct}{\mcitedefaultseppunct}\relax
\EndOfBibitem
\bibitem[Gao \latin{et~al.}(2016)Gao, Gong, Titze, Almeida, Ajayan, and Li]{34}
Gao,~F.; Gong,~Y.; Titze,~M.; Almeida,~R.; Ajayan,~P.~M.; Li,~H. Valley trion
  dynamics in monolayer ${\mathrm{MoSe}}_{2}$. \emph{Phys. Rev. B}
  \textbf{2016}, \emph{94}, 245413\relax
\mciteBstWouldAddEndPuncttrue
\mciteSetBstMidEndSepPunct{\mcitedefaultmidpunct}
{\mcitedefaultendpunct}{\mcitedefaultseppunct}\relax
\EndOfBibitem
\bibitem[sup()]{supl}
Supplementary Information. \relax
\mciteBstWouldAddEndPunctfalse
\mciteSetBstMidEndSepPunct{\mcitedefaultmidpunct}
{}{\mcitedefaultseppunct}\relax
\EndOfBibitem
\bibitem[Yu \latin{et~al.}(2018)Yu, Liu, and Yao]{br}
Yu,~H.; Liu,~G.-B.; Yao,~W. Brightened spin-triplet interlayer excitons and
  optical selection rules in van der Waals heterobilayers. \emph{2D Materials}
  \textbf{2018}, \emph{5}, 035021\relax
\mciteBstWouldAddEndPuncttrue
\mciteSetBstMidEndSepPunct{\mcitedefaultmidpunct}
{\mcitedefaultendpunct}{\mcitedefaultseppunct}\relax
\EndOfBibitem
\bibitem[Joe \latin{et~al.}(2021)Joe, Jauregui, Pistunova, Mier~Valdivia, Lu,
  Wild, Scuri, De~Greve, Gelly, Zhou, Sung, Sushko, Taniguchi, Watanabe,
  Smirnov, Lukin, Park, and Kim]{ec}
Joe,~A.~Y. \latin{et~al.}  Electrically controlled emission from singlet and
  triplet exciton species in atomically thin light-emitting diodes. \emph{Phys.
  Rev. B} \textbf{2021}, \emph{103}, L161411\relax
\mciteBstWouldAddEndPuncttrue
\mciteSetBstMidEndSepPunct{\mcitedefaultmidpunct}
{\mcitedefaultendpunct}{\mcitedefaultseppunct}\relax
\EndOfBibitem
\bibitem[Rigosi \latin{et~al.}(2015)Rigosi, Hill, Li, Chernikov, and
  Heinz]{RN12}
Rigosi,~A.~F.; Hill,~H.~M.; Li,~Y.; Chernikov,~A.; Heinz,~T.~F. Probing
  Interlayer Interactions in Transition Metal Dichalcogenide Heterostructures
  by Optical Spectroscopy: MoS2/WS2 and MoSe2/WSe2. \emph{Nano Letters}
  \textbf{2015}, \emph{15}, 5033--5038\relax
\mciteBstWouldAddEndPuncttrue
\mciteSetBstMidEndSepPunct{\mcitedefaultmidpunct}
{\mcitedefaultendpunct}{\mcitedefaultseppunct}\relax
\EndOfBibitem
\bibitem[Singh \latin{et~al.}(2016)Singh, Moody, Tran, Scott, Overbeck,
  Bergh\"auser, Schaibley, Seifert, Pleskot, Gabor, Yan, Mandrus, Richter,
  Malic, Xu, and Li]{38}
Singh,~A. \latin{et~al.}  Trion formation dynamics in monolayer transition
  metal dichalcogenides. \emph{Phys. Rev. B} \textbf{2016}, \emph{93},
  041401\relax
\mciteBstWouldAddEndPuncttrue
\mciteSetBstMidEndSepPunct{\mcitedefaultmidpunct}
{\mcitedefaultendpunct}{\mcitedefaultseppunct}\relax
\EndOfBibitem
\bibitem[Karni \latin{et~al.}(2022)Karni, Barré, Pareek, Georgaras, Man,
  Sahoo, Bacon, Zhu, Ribeiro, O’Beirne, Hu, Al-Mahboob, Abdelrasoul, Chan,
  Karmakar, Winchester, Kim, Watanabe, Taniguchi, Barmak, Madéo, da~Jornada,
  Heinz, and Dani]{39}
Karni,~O. \latin{et~al.}  Structure of the moiré exciton captured by imaging
  its electron and hole. \emph{Nature} \textbf{2022}, \emph{603},
  247--252\relax
\mciteBstWouldAddEndPuncttrue
\mciteSetBstMidEndSepPunct{\mcitedefaultmidpunct}
{\mcitedefaultendpunct}{\mcitedefaultseppunct}\relax
\EndOfBibitem
\bibitem[Baek \latin{et~al.}(2021)Baek, Brotons-Gisbert, Campbell, Vitale,
  Lischner, Watanabe, Taniguchi, and Gerardot]{RN30}
Baek,~H.; Brotons-Gisbert,~M.; Campbell,~A.; Vitale,~V.; Lischner,~J.;
  Watanabe,~K.; Taniguchi,~T.; Gerardot,~B.~D. Optical read-out of Coulomb
  staircases in a moiré superlattice via trapped interlayer trions.
  \emph{Nature Nanotechnology} \textbf{2021}, \emph{16}, 1237--1243\relax
\mciteBstWouldAddEndPuncttrue
\mciteSetBstMidEndSepPunct{\mcitedefaultmidpunct}
{\mcitedefaultendpunct}{\mcitedefaultseppunct}\relax
\EndOfBibitem
\end{mcitethebibliography}

\end{document}